\documentstyle[onecolumn,epsfig]{mn}
\newif\ifAMStwofonts

\newcommand{\lsim}{\,\lower2truept\hbox{${<\atop\hbox{\raise4truept\hbox{$\sim$}}}$}\,}
\newcommand{\gsim}{\,\lower2truept\hbox{${>\atop\hbox{\raise4truept\hbox{$\sim$}}}$}\,}

\def\lsim{\,\lower2truept\hbox{${< \atop\hbox{\raise4truept\hbox{$\sim$}}}$}\,}
\def\gsim{\,\lower2truept\hbox{${> \atop\hbox{\raise4truept\hbox{$\sim$}}}$}\,}

\def\deg{\ifmmode^\circ \else$^\circ $\fi}    
\def\arcs{\ifmmode {'' }\else $'' $\fi}     
\def\arcm{\ifmmode {' }\else $' $\fi}     
\def\buildrel#1\over#2{\mathrel{\mathop{\null#2}\limits^{#1}}}
\def\mper{\ifmmode \buildrel m\over . \else $\buildrel m\over .$\fi}
\def\hper{\ifmmode \rlap.^{h}\else $\rlap{.}^h$\fi}
\def\sper{\ifmmode \rlap.^{s}\else $\rlap{.}^s$\fi}
\def\arcsper{\ifmmode \rlap.{' }\else $\rlap{.}' $\fi}
\def\arcmper{\ifmmode \rlap.{'' }\else $\rlap{.}'' $\fi}

\def\mincir{\ \raise -2.truept\hbox{\rlap{\hbox{$\sim$}}\raise5.truept	
\hbox{$<$}\ }}								%
\def\magcir{\ \raise -2.truept\hbox{\rlap{\hbox{$\sim$}}\raise5.truept	%
\hbox{$>$}\ }}								%

\title[WMAP and the Generalized Chaplygin Gas] 
{WMAP and the Generalized Chaplygin Gas} 

\author[L. Amendola et al.]
{L. Amendola$^1$,
F. Finelli$^{2\,*}$, C. Burigana$^2$, and D. Carturan$^2$ 
\\ 
$^1$INAF/Osservatorio Astronomico di Roma, Via Frascati 33, I-00040
Monte Porzio Catone, Italy \\
$^2$IASF/CNR, Istituto di Astrofisica Spaziale e Fisica Cosmica,
Sezione di Bologna, \\
Consiglio Nazionale delle Ricerche, 
Via Gobetti 101, I-40129 Bologna, Italy}

\date{Submitted to MNRAS, 5 May 2003.}

\pagerange{\pageref{firstpage}--\pageref{lastpage}}
\pubyear{2003}

\begin{document}

\maketitle

\label{firstpage}
\footnotetext{E-mail: amendola@coma.mporzio.astro.it ; name@bo.iasf.cnr.it}
\footnotetext{$^*$ Also supported by INFN, Sezione di Bologna, via 
Irnerio, 46 -- I-40126 Bologna -- Italy}
\begin{abstract}
We compare the WMAP temperature power spectrum and SNIa data to models
with a generalized Chaplygin gas as dark energy. The generalized Chaplygin
gas is a component with an exotic equation of state, 
\( p_{X}=-A/\rho ^{\alpha }_{X} \)
(a polytropic gas with negative constant and exponent). Our main result
is that, restricting to a flat universe and to adiabatic pressure
perturbations for the generalized Chaplygin gas, the constraints at
95\% CL to the present equation of state \( w_{X} =  p_{X} / \rho_{X}\) 
and to the parameter \( \alpha  \) are \( -1\leq w_{X}<-0.8 \), \( 0\leq 
\alpha <0.2 \), respectively. Moreover, we show that a Chaplygin gas (\( 
\alpha =1 \))
as a candidate for dark energy is ruled out by our analysis at more
than the \( 99.99\% \) CL. 
A generalized Chaplygin gas as a unified dark matter candidate 
\( (\Omega_{CDM}=0) \)
appears much less likely than as a dark energy model,
although its \( \chi^2 \) is only two sigma away from the expected value.
\end{abstract}


\section{\textbf{Introduction}}

The picture of a universe mostly filled by an unknown component, Dark
Energy (DE henceforth), seems overwhelmingly justified by the latest
observations of CMB anisotropies (Spergel et al. 2003). 

While there is a strong degeneracy of the multitude of DE candidates
for the background evolution --~and therefore in the SNIa data 
(Perlmutter et al. 1999; Riess et al. 1998)~--, 
additional information should be
contained in the linear fluctuations. However, in order to differentiate
among DE models, the avenue of studying DE fluctuations is not promising
as it seems. Infact, in the simplest uncoupled quintessence models
(Ratra \& Peebles 1988; Frieman et al. 1995; Caldwell, Dave \& Steinhardt
1995) based on scalar fields with ordinary Klein-Gordon Lagrangian,
quintessence fluctuations are evanescent on sub-horizon scales. This
is due to the presence of a Jeans scale for scalar field perturbations,
of the order of the Hubble radius. This Jeans scale is related to
a unitary sound speed (in \( c=1 \) units) for scalar field perturbations. 
For the same reason, isocurvature perturbations caused by an offset of
the quintessence component (Abramo \& Finelli 2001) are not observationally
harmful as other isocurvature modes. 

A possibility of having a smaller Jeans scale for dark energy fluctuations
is offered by theories with non standard kinetic terms for the scalar
field candidate. A scalar field with a Born-Infeld action is an example:
\begin{equation}
\label{action}
S=\int d^{4}x\sqrt{-g}\, V\, \sqrt{1+\frac{\partial _{\mu }\varphi 
\partial ^{\mu }\varphi }{M^{4}}}\, ,
\end{equation}
where \( M \) is a fundamental scale and \( V \) is a potential. 
Recently, the tachyon field
(with \( V=V(\varphi ) \)) motivated from string theory has received
a lot of attention (Sen 2002, Gibbons 2002, Padmanabhan 2002). A Chaplygin
gas is a hydrodinamical description of the scalar field with the Born-Infeld
Lagrangian in Eq. (\ref{action}) with constant \( V \) (Kamenshchik, 
Moschella \& Pasquier
2001). A Chaplygin gas is characterized by a pressure which is inversely
proportional to the energy \( p\propto -1/\rho  \). A generalized
Chaplygin gas (henceforth GCG) is a perfect fluid with a polytropic
equation of state (Bento, Bertolami \& Sen 2002): \begin{equation}
\label{pressure}
p_{X}=-\frac{A}{\rho _{X}^{\alpha }}
\end{equation}
 where \( \alpha  \) is a parameter between \( 0 \) and \( 1 \)
in order to have a sound speed at most luminal for perturbations.
\( A \) is a constant with dimensions \( [M^{4(1+\alpha )}] \).
A GCG is a phenomenological extension of the Chaplygin gas, which
exhausts all the possibilities for a polytropic perfect fluid dark
energy candidate, whose perturbations are stable on small scales (Fabris,
Goncalves \& Souza 2002a, Carturan \& Finelli 2002 --~henceforth CF~--). 

The behaviour in time of a GCG interpolates between dust and a cosmological
constant, with an intermediate behaviour as \( p=\alpha \rho  \).
The Jeans instability of GCG perturbations is first similar to CDM
fluctuations (when the GCG has a negligible pressure) and then disappears
(when the GCG behaves as a cosmological constant) (see CF). 
Both this late suppression of GCG fluctuations and the apperance of a non 
zero Jeans length leave a large integrated Sachs-Wolfe (ISW) imprint on 
the CMB anisotropies (CF). A model in 
which the GCG is the only dark component is referred to as a unified dark
matter/dark energy (UDM) model. 

The viability of GCG as DE and UDM model has been first analyzed in
the context of Supernovae (Fabris et al. 2002b;
Avelino et al. 2002; Makler, de Oliveira \& Waga 2003) and other 
complementary data (Dev, Jain \& Alcaniz 2003, Alcaniz, Jain and Dev 
2003). The comparison
with CMB data has been the next step (CF, Bean \& Dore' 2003) and
it has led to much stronger constraints (Bean \& Dore' 2003). 

The imprint of a GCG is also present on the matter power spectrum, since GCG 
perturbations affect both CMB anisotropies and structure formation. This 
last issue has been considered by Sandvik et al. (2002) in the context of 
UDM models, claiming that a GCG is ruled out as a UDM candidate at \( 
99.999\% \) (Sandvik et al. 2002). However, their analysis does not take 
into account baryons, while we will show they are relevant for the shape 
of the total matter power spectrum.
Moreover, the analysis by Sandvik et al. is based
on a linear treatment of perturbations until the present time, neglecting  
any non linear effects which may be important and unexpected 
because of the time dependence of the GCG Jeans length. Such nonlinear 
effects should be more important for the present matter
power spectrum than for the CMB spectrum. For this reason, we focus
on the comparison with CMB data.

In this paper we compare the predictions of GCG models with the recently 
released WMAP data, which provide the most precise determination of the 
CMB spectrum so far (Hinshaw et al. 2003), and with SNIa data (Perlmutter 
et al. 1999, Riess et al. 1998).

\section{The Model}

In this section we review a brief and basic description of the model
(for more details see CF). The conservation equation for the GCG in
a Robertson-Walker metric is solved by \begin{equation}
\label{rhox}
\rho _{X}=[A+\frac{B}{a^{3(\alpha +1)}}]^{\frac{1}{1+\alpha }},
\end{equation}
 where \( a \) is the scale factor ($a=1$ today) and \( B \) is an 
integration
constant with dimension \( [M^{4(1+\alpha )}] \). 

The equation of state is therefore \begin{equation}
\label{wx}
w_{X}(a)=\frac{p_{X}}{\rho _{X}}=-\frac{A}{\rho _{X}^{1+\alpha }} \,,
\end{equation}
while the present equation of state is \[
w_{X}=-\frac{A}{A+B}.\]
Instead of \( A,B \) we find convenient to employ the parameters
\( \Omega _{X},w_{X} \) , where \( \Omega _{X} \) is the present
value of the density parameter of the GCG and\begin{eqnarray*}
A & = & -w_{X}(\Omega _{X}\rho _{c})^{1+\alpha },\\
B & = & (1+w_{X})(\Omega _{X}\rho _{c})^{1+\alpha },
\end{eqnarray*}
where \( \rho _{c} \) is the present critical density. Beside the
GCG we introduce for generality also baryons and CDM. Although the
GCG is very interesting as an unified model of dark energy and dark 
matter, it is conceivable that it is in fact only a DE component 
additional to ordinary CDM and baryons. 

Perturbations of GCG are stable on small scales since the sound speed
is positive (\( c_{X}^{2}=\partial p_{X}/\partial \rho _{X}=-\alpha w_{X} \)).
An important consequence of assuming a perfect fluid instead of a
scalar field as a candidate of dark energy is the absence of intrinsic
non adiabatic pressure perturbations. Infact, for a GCG pressure perturbations
are locked to density perturbations \begin{equation}
\label{locking}
\delta p_{X}=c_{X}^{2}\delta \rho _{X}\, .
\end{equation}
Instead, in presence of non adiabatic pressure perturbations, we
have in general: \begin{equation}
\delta p_{X}=c_{X}^{2}\delta \rho _{X}+\Delta p_{X}\, .
\end{equation}
For example, in a scalar field model with standard kinetic term for
quintessence \( c_{X}^{2}=1 \) and \( \Delta p_{X}=2V_{\phi }\delta \phi  \),
where \( V_{\phi } \) is the first derivative of the potential and
\( \delta \phi  \) is the field fluctuation. 

A perfect fluid description of a GCG offers the chance to explore
the possibility of the basic relation (\ref{locking}) for dark energy
pressure perturbations. 

In Figs. 1 and 2 we compare the \( C_{\ell } \) temperature spectrum
with the WMAP data varying \( w_{X} \) and \( \alpha  \), respectively
(see also CF). These figures show how strongly the \( C_{\ell } \) 
spectrum depends on perturbations of GCG, when this is a DE candidate
(\( \Omega _{c} \) is fixed to \( 0.27 \)). Fig. 1 shows how the
resulting CMB spectrum strongly differ even for models which have
\( w_{X}\leq -0.91 \). Notice that the 2nd and 3rd peak move to the
left increasing \( w_{X} \), because then the expansion is less and
less accelerated, while the 1st peak remains more or less stable because
of the concurring strong integrated Sachs-Wolfe (ISW) effect. Fig. 2 shows 
the strong dependence
of CMB spectrum on \( \alpha  \), which regulates the Jeans length
of GCG perturbations and therefore the amount of ISW effect. 
The peak-to-plateau ratio decreases for \( \alpha  \) increasing,
because after the dust behaviour the GCG induces a strong ISW. 
The dependence of CMB
spectrum and LSS on the Jeans length of DE perturbations is a generic
effect and it has been also found recently in K-essence models (DeDeo,
Caldwell \& Steinhardt 2003), but it is less severe than in GCG
models.

\begin{figure}
{\centering \resizebox*{!}{12cm}{\includegraphics{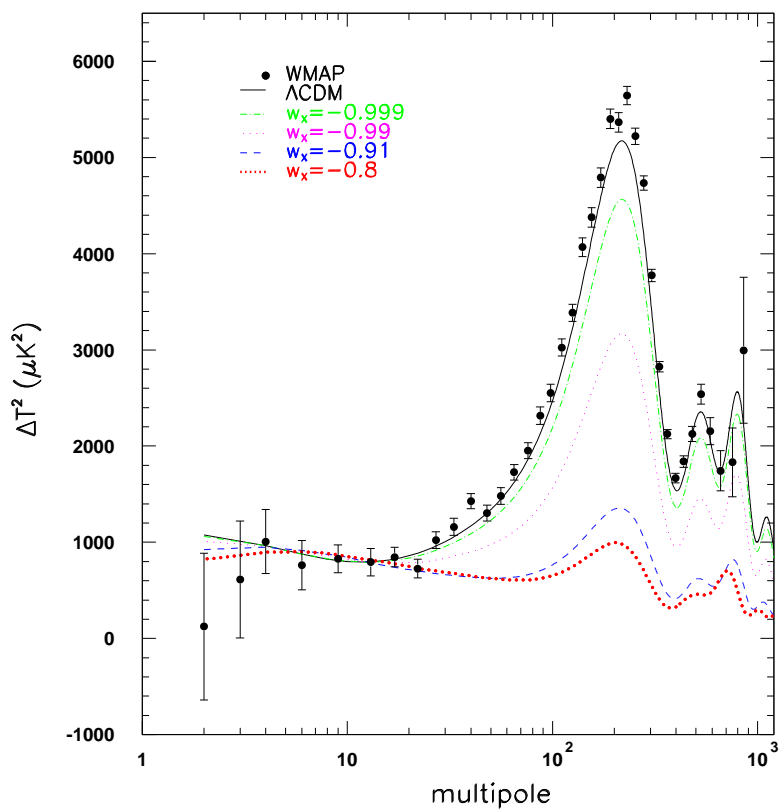}} \par}

\caption{The \protect\protect\( C_{\ell }\protect \protect \) spectrum varying
\protect\protect\( w_{X}\protect \protect \) is plotted in comparison
with a \protect\( \Lambda \protect \)CDM model (the upper curve):
\protect\( w_{X}\simeq -0.8,-0.91,-0.99,-0.999\protect \) from bottom
to top, respectively. The other parameters are \protect\( \alpha 
=1\protect \),
\protect\( h=0.72\protect \), \protect\( w_{b}=0.024\protect \),
\protect\( w_{c}=0.14\protect \), \protect\( n_{s}=1\protect \).
The spectra are COBE normalized at \protect\( \ell =10\protect \).}
\end{figure}

\begin{figure}
{\centering \resizebox*{!}{12cm}{\includegraphics{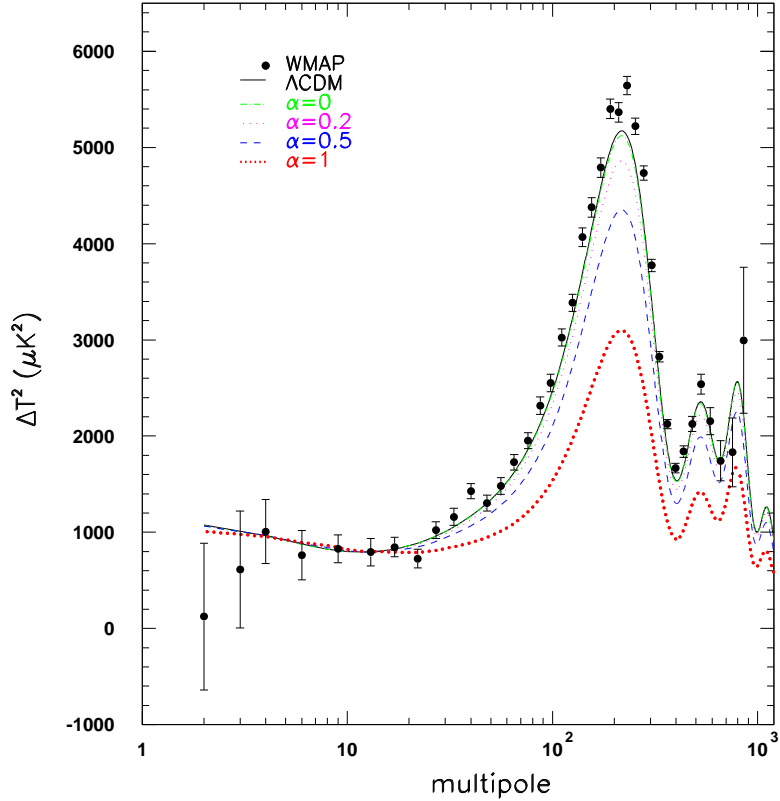}} \par}

\caption{The \protect\protect\( C_{\ell }\protect \protect \) spectrum varying
\protect\protect\( \alpha \protect \protect \) is plotted in comparison
with a \protect\( \Lambda \protect \)CDM model (the upper solid line):
\protect\( \alpha =1,0.5,0.2,0\protect \) from bottom to top, respectively.
The other parameters are \protect\( w_{X}\simeq -0.99\protect \),
\protect\( h=0.72\protect \), \protect\( w_{b}=0.024\protect \),
\protect\( w_{c}=0.14\protect \), \protect\( n_{s}=1\protect \).
The spectra are COBE normalized at \protect\( \ell =10\protect \).
We note how the \protect\( \Lambda \protect \)CDM curve and
\protect\( \alpha =0\protect \) are very close, but not identical
because of the different perturbation sectors.}
\end{figure}

\section{Likelihood analysis }

We compare the models to the combined power spectrum estimated by
WMAP (Hinshaw et al. 2003). To derive the likelihood we adopt a version
of the routine described in Verde et al. (2003), which takes into
account all the relevant experimental properties (calibration, beam
uncertainties, window functions, etc). Since the likelihood routine
employs approximations that work only for spectra not too far from
the data, we run it only for models whose \( \chi ^{2} \) per d.o.f. is 
less than 1.5 from the number of degrees of freedom. 

Our theoretical model depends on two GCG parameters, four
cosmological parameters and the overall normalization \( N \): 
\begin{equation}
\alpha ,w_{X},n_{s},h,\omega _{b},\omega _{c},N.
\end{equation}
where \( \omega _{b,c}=\Omega _{b,c}h^{2} \) and $n_s$ is the primordial 
fluctuation slope. As anticipated, to
save computing time we found necessary to restrict the analysis
to a flat space; moreover, we fixed the optical depth to \( \tau =0.17 \),
the mean value found by WMAP (Spergel et al. 2003). The overall normalization
has been integrated out numerically. We calculate the theoretical
\( C_{\ell ,t} \) spectra by a modified parallelized CMBFAST (Seljak
\& Zaldarriaga 1996) code that includes the full set of perturbation
equations (see CF). We do not include gravitational waves and the
other parameters are set as follows: \( T_{cmb}=2.726K, \) 
\( Y_{He}=0.24,N_{\nu }=3.04 \). 

We evaluated the likelihood on a non-uniformly spaced grid of 
roughly \( 50,000 \) models
(for each normalization) with the following top-hat broad priors:
\( w_{X}\in (-1,-0.5), \) \( \quad \alpha \in (0,.5),\quad  \)\ \( n_{s}\in (0.8,1.2), \)
\( \quad \omega _{b}\in (0.005,0.04),\quad  \)\ \( \omega _{c}\in (0.0,0.2) \). 
For the Hubble constant we adopted the top-hat prior \( h\in (0.5,0.9); \)
we also employed the HST result (Freedman et al. 2001) \( h=0.72\pm 0.08 \)
(Gaussian prior). One reason to adopt a grid approach than a Markov chain 
method (as in Verde et al. 2003) is that the independent grid evaluation 
can be parallelized with maximum efficiency.

\section{Results from comparison with WMAP}

In Fig. 3 we plot the likelihood functions for each parameter, marginalizing
in turn over the others. The horizontal lines mark the 68\% and 95\%
CL; the vertical lines the 68\% and 95\% upper bounds. We obtain
the bounds \begin{eqnarray}
\alpha  & < & 0.2(0.05)\nonumber \\
w_{x} & < & -0.8(-0.92)\label{results} 
\end{eqnarray}
 at 95\%(68\%) CL. These constraints are the most stringent ones
obtained so far using only CMB data; the limit on \( \alpha  \) is
5--10 stronger than the previous one obtained with pre-WMAP data (Bean
\& Dore' 2002). For the other parameters we obtain results not very
different than the standard ones. We also show the results imposing
the HST prior \( h=0.72\pm 0.08 \); they are almost identical to
the one without prior on \( h \). The best fit DE model is $n_s=1$, 
$\alpha=0$, $w_X=-0.98$, $h=0.71$, $\omega_b=0.024$, $\omega_c=0.12$ has  
$\chi^2$/d.o.f.=979/892.

\begin{figure}
{\centering \resizebox*{!}{12cm}{\includegraphics{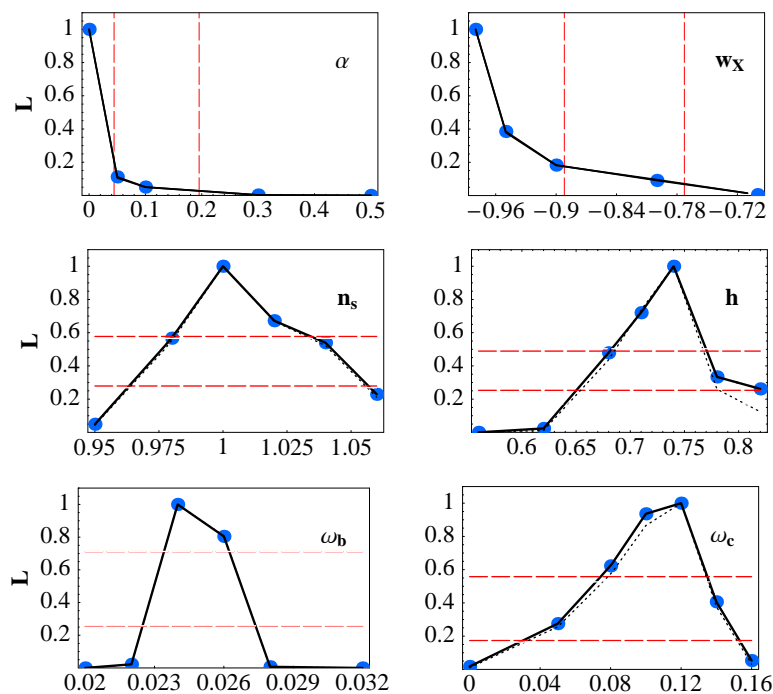}} \par}

\caption{Marginalized likelihood functions for the six cosmological parameters.
The dotted lines are for the HST prior on the Hubble constant. The
horizontal long-dashed lines are the confidence levels at 68\% and
95\%. The vertical long-dashed lines in the panels for 
\protect\protect\( \alpha \,, w_{X}\protect \protect \)
mark the upper bounds at 68\% and 95\% CL.}
\end{figure}

\begin{figure}
{\centering \resizebox*{!}{12cm}{\includegraphics{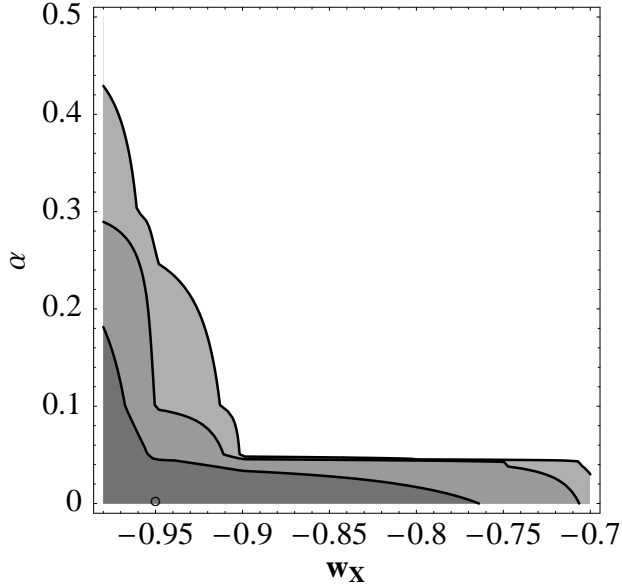}} \par}

\caption{Likelihood in the plane 
\protect\protect\( w_{X},\alpha \protect \protect \).
The contours are at the 68\%, 95\% and 99\% CL, inside to outside.}
\end{figure}

\section{Constraints on the Unified Dark Model }

Any DE model which interpolates in time between a dust epoch and an
accelerating stage has the possibility of playing the role of a UDM
candidate, if one renounces to the presence of CDM. Naturally, in
such a case one should explain also why 30\% or so of the matter collapsed
into galaxies and clusters while the remaining did not. The GCG has
received a lot of attention also in this light of a UDM model. While
SNIa data allow for this possibility, it was claimed that LSS data
completely ruled it out (Sandvik et al. 2002). Here we find that WMAP data 
alone (for which a linear analysis is sufficient)
show that a GCG as a unified candidate is much less likely than a model 
which includes a CDM component. The likelihood for $\omega_c=0$ is 
infact 
roughly 50 times
smaller than for $\omega_c=0.12$; formally, this rules out the UDM at 
slightly more that 99.99\% C.L. when compared with GCG as DE. The lowest 
value allowed by WMAP data is 0.1 (95%
C.L.). However, if we calculate the $\chi^2$ for the best fit among the 
UDM cases (corresponding
to a model with $n_s=0.98$, $\alpha=0$, $w_X=-0.8$, $h=0.82$, 
$\omega_b=0.026$) we 
find $\chi^2$/d.o.f.=986/893, indicating that the UDM model is an 
acceptable fit to WMAP data at at 2$\sigma$ level.


Note that our results have been obtained varying also $n_s$, which has 
been instead fixed in the analysis of Bean \& Dore' (2002) and of Sandvik 
et al. (2002) (who also fixed $h$).

\begin{figure}
{\centering \resizebox*{!}{12cm}{\includegraphics{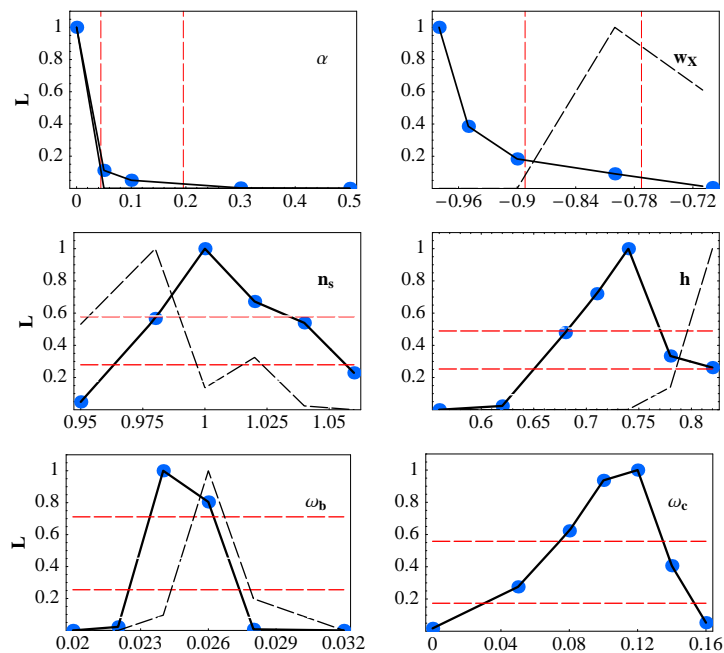}} \par}

\caption{Likelihood functions for the GCG model (full curve) and the unified
model (\protect\protect\( \Omega _{c}=0\protect \protect \) ; dashed
curve). The horizontal long-dashed lines are the confidence levels
at 68\% and 95\%. The vertical long-dashed lines in the panels for
\protect\protect\( \alpha \,, w_{X }\protect \protect \) mark the
upper bounds at 68\% and 95\% CL.}
\end{figure}

\section{Adding the Supernovae constraints}

We now proceed with a further refinement of our result, by taking
into account the constraint from the SN I a Hubble diagram. This has
been evaluated by several groups (Fabris et al. 2002a;
Avelino et al. 2002; Makler et al. 2003), but 
not in a suitable form for our purposes. 

The luminosity distance in a GCG model in flat space is \[
d_{L}(z)=(1+z) \int _{0}^{z}\frac{dz'}{H(z')},\]
where the Hubble function is
\[
H^{2}(z)=H_{0}^{2}\left[ \Omega _{m}(1+z)^{3}+(1-\Omega _{m})
((1+w_{X})(1+z)^{3(1+\alpha )}-w_{X})^{\frac{1}{1+\alpha }}\right] ,\]
 and \( \Omega _{m}=\Omega _{b}+\Omega _{c} \). 

We compare the luminosity distance to the SCP data of Perlmutter et
al. (1999) (their fit C), to which we add the supernova at \( z\approx 1.7 \)
(Benitez et al. 2002). We show in Fig. 6 the 2-dimensional likelihood
function marginalized over the {}``nuisance parameters{}'' (see
definitions in Perlmutter et al. 1999) \( \alpha ,M^{*} \) , and
over \( \Omega _{m} \) with Gaussian prior \( \Omega _{m}=0.3\pm 0.1 \).
Notice that the contours are almost insensitive to \( \alpha \,  \) since \( d_{L}(z) \)
at small redshifts is independent of \( \alpha  \). 
This shows how CMB data constrain the GCG models much more than SNIa 
data. In Fig. 7 we
multiply the SN and WMAP likelihood functions. The final constraints
turn out to be almost identical to those in (\ref{results}).

\begin{figure}
{\centering \resizebox*{!}{12cm}{\includegraphics{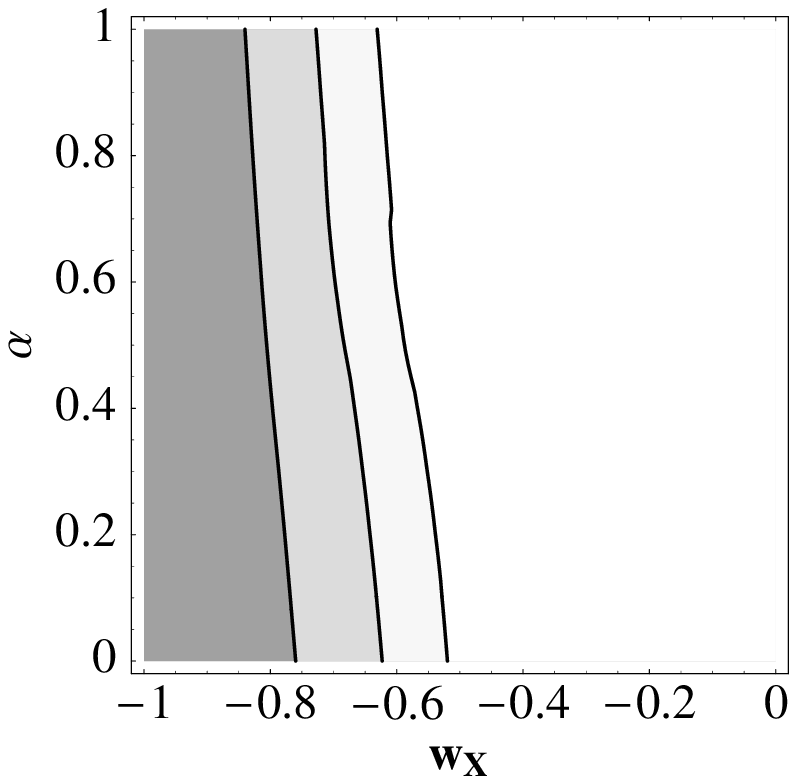}} \par}

\caption{SNIa likelihood in the plane \protect\protect\( w_{X} \,, 
\alpha \protect \protect \),
with a Gaussian prior \protect\protect\( \Omega _{m}=0.3\pm 0.1\protect \protect \).
Confidence regions at 65\%, 95\% and 99\% from dark to light gray.}
\end{figure}

\begin{figure}
{\centering \resizebox*{!}{12cm}{\includegraphics{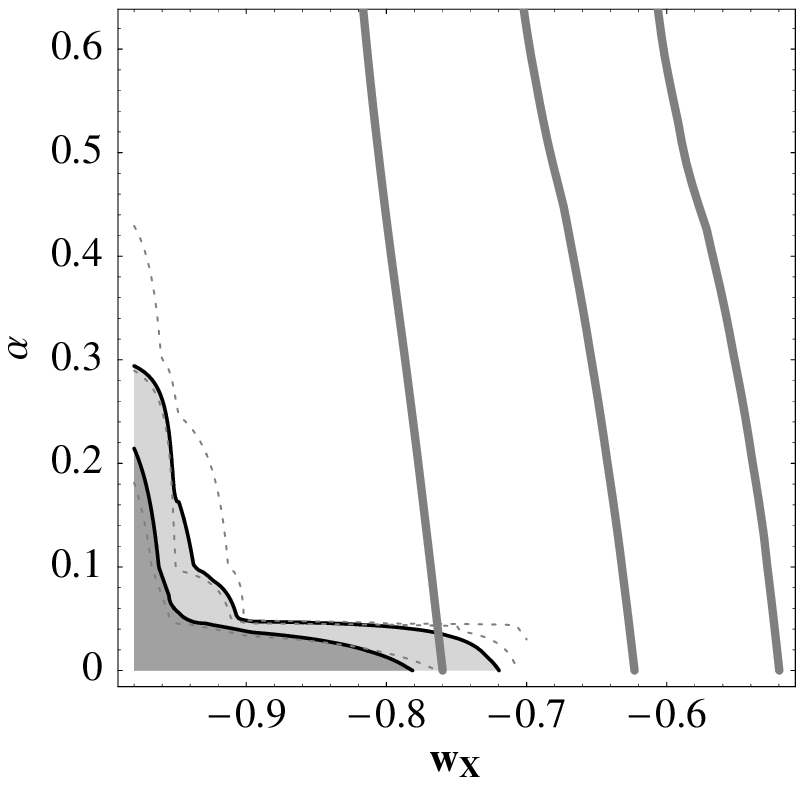}} \par}

\caption{Adding the SNIa constraint. The thick gray curves are the confidence
regions for the SNIa; the dotted curves the confidence regions for
the CMB; the gray regions represent the combined likelihood (65\%
and 95\% c.l.). As it can be seen, the SNIa constraints add little
information.}
\end{figure}

\section{Considerations on the Mass Power Spectrum }

The influence of the GCG on structure formation is another important
observational imprint of this candidate for DE and UDM model. GCG
fluctuations accumulate on all scales when the GCG behaves as CDM.
As soon as the GCG equation of state drops below zero, the GCG fluctuations
are no longer Jeans unstable at all scales, oscillating and decaying. This 
exotic behaviour of fluctuations occurs also when the GCG simulates \( 
\Lambda  \). As already said in the introduction, this suppression of 
fluctuations is in part responsible for the large ISW effect in CMB 
anisotropies. It is clear that a similar behaviour of fluctuations differs 
both from CDM and standard quintessence, and it is more important in 
absence of CDM, i.e. in a UDM model.

Sandvik et al. (2002) used a comparison of the predicted power spectrum in 
GCG models with LSS data --~2dF survey (Colless et al. 2001; see also 
Tegmark, Hamilton \& Xu 2002)~-- to rule out 
the GCG as a UDM candidate at 99.999\% CL. Their approximate analysis 
neglects baryons (this has also been noted recently by Beca et al. 
2003) and is based
on a linear treatment of perturbations until the present time. However,
the nonlinear clustering of GCG could be rather nontrivial because
of the time dependence of its Jeans length and deserves more study.

In Fig. 8 we compare the predicted power spectra --~total (solid line), 
GCG (dashed line) and baryon (dotted line)~-- for UDM model
with the APM (Padilla \& Baugh 2003) and 2dF (Colless et al. 2001) data 
when $\alpha$ varies.
The power spectra are obtained through the transfer functions of the 
modified CMBFAST code (CF), i.e. by using the linear evolution. We do not 
show as target the CDM power 
spectrum because almost indistinguishable from the baryon one.
The power spectra are normalized 
at $k=0.01$ h/Mpc as in Sandvik et al. (2002) and the results in 
Fig. 8 can be directly compared with those of Fig. 1 in Sandvik et al.. 
It is very interesting to note the role of baryons, which were neglected 
by Sandvik et al.: baryons keep on clustering on all scales, at all time 
after decoupling. Even if baryons are a subdominant component ($\Omega_b = 
0.05$), they are very important when one considers the total matter power 
spectrum $P_T (k) = \Omega_b^2 P_b (k) + \Omega_X^2 P_X (k)$. 
The inclusion of baryons smooths out the oscillations of the GCG component 
in the total matter power spectrum, in particular for $\alpha \ge 
10^{-3}$: when comparing to the LSS data, 
only the shape is relevant and not the features of the GCG component.
We also note that for $\alpha \ge
10^{-2}$ the appearance of a finite Jeans length 
and/or of a suppression of DE component maybe very interesting in connection 
with the bump present in APM data. Therefore, it appears that an 
intermediate value 
of $\alpha$ ($0.01 \le \alpha \le 0.2$) may fit both CMB and LSS data.

On concluding, in UDM models the comparison of the total matter power 
spectrum with LSS changes by including baryons, even within 
the linear approximation. In the context of DE models, this effect is even 
more important since CDM perturbations clustering on all scales are also 
present.

\begin{figure}
{\centering \resizebox*{!}{12cm}{\includegraphics{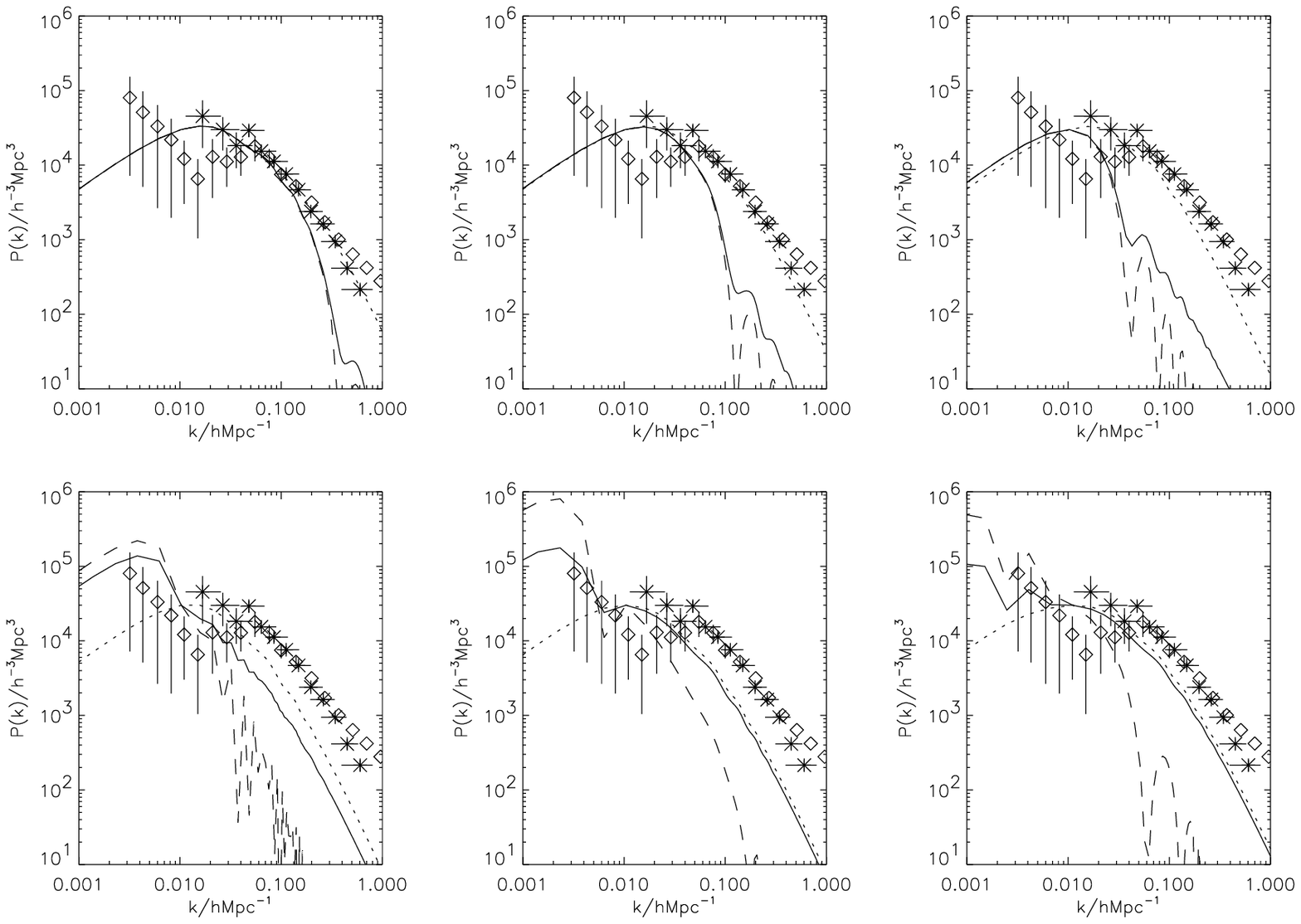}} \par}

\caption{We compare the predictions of a UDM model based on GCG with 
APM (diamonds)
and 2dF (asterisks) surveys when $\alpha$ varies ($\alpha = 10^{-5}, 
10^{-4}, 10^{-3}, 
10^{-2}, 0.05, 0.25$ from left to right, top to bottom).  
The other parameters are $h=0.7$, $w_X \simeq - 0.74$, $\Omega_X = 0.95$ 
and $\Omega_b = 0.05$.}
\end{figure}

\section{Conclusions}

We have analyzed quantitatively the observational effects of a GCG
as a DE candidate. We have restricted our analysis to a flat universe
and to purely adiabatic pressure perturbations. 

So far, most authors studied only the comparison of the Hubble law 
predicted in 
presence of a GCG with the SNIa (Fabris et al. 2002a; 
Avelino et al. 2002; Makler et al. 2003). However, 
as also our Figs. 6--7 show, 
SNIa data constrain GCG models very weakly. For instance,
the Hubble law is almost insensitive to \( \alpha  \), as we see in 
Fig. 6. Instead, \( \alpha  \) is the maximum value which the sound
speed can reach and it is related to the time at which the Jeans 
instability disappears and therefore fluctuations are very sensitive to 
this quantity:
this results in a strong dependence of the CMB spectrum on \( \alpha  \),
as we see from Fig. 2. 

The combined result of CMB and SNIa data is presented in Fig. 7, which
leads to \( \alpha < 0.2 \) and \( w_{X}<-0.8 \) at \( 95\% \) c.l.
The bound imposed by WMAP data on \( \alpha  \) is 5--10 stronger
than the previous data (Bean \& Dore 2003), and the Chaplygin gas
(\( \alpha =1 \)) is ruled out as a DE candidate at more than the
\( 99.99\% \) c.l. 

The possibility of a unified dark model with a GCG (corresponding
to \( \Omega _{c}=0 \)) seems disfavoured by the latest CMB anisotropy
data in comparison to the GCG playing the role of a DE model, although 
the UDM on its own is not a bad fit to the data. The 
$\chi^2$ statistics is 987/893 for a UDM model, whereas is 979/892 for a 
DE model.
Our result is in contrast with the analysis
of the peaks position recently performed by Bento, Bertolami and Sen
(2003), in which a different region in the (\( \alpha \, ,w_{X} \))
plane is allowed. This disagreement may come from an inaccurate analytic
estimate of the peaks position (already pointed out by CF). 

We have also addressed the problem of the comparison with the LSS data. 
We have compared results on the matter power spectrum with 2dF and APM 
data, within the linear approximation. By considering the worst case, 
i.e. a UDM model, we find that LSS data may be complementary to CMB data. 
The main question about the validity of linear treatment on such short 
scales however remains.

On concluding, standard quintessence models seem the most economic
variation of DE models with respect to \( \Lambda  \)CDM. Differentiating
among DE models by studying the sound speed of its perturbations seems
a promising avenue and this has already been shown in two different
class of models (CF; DeDeo et al. 2003). We have
shown that GCG models are models with high predictive power when compared
with CMB anisotropies. Current and 
future~\footnote{http://astro.estec.esa.nl/Planck} 
high precision CMB anisotropy data can finally constrain and/or 
rule out physical scenarios based on different models of DE.

\section*{Acknowledgements}

Most of the computations have been performed on a 128-processors Linux
cluster at CINECA. We thank the staff at CINECA for support.
F. F. would like to thank R. Abramo for many discussions on this topic.
L. A. and F. F. would like to thank O. Bertolami and I. Waga for useful 
discussions.

\end{document}